# Rendezfood: A Design Case Study of a Conversational Location-based Approach in Restaurants

Rendezfood: Designing Conversational Dining Experiences

Philip Weber, EduTech for Digital Transformation, University of Hagen, Hagen, Germany,
ORCID: 0000-0003-3537-5753

Kevin Krings, Information Systems, University of Siegen, Siegen, Germany,
ORCID: 0000-0002-8772-6132

Lukas Schröder, EduTech for Digital Transformation, University of Hagen, Hagen, Germany
ORCID: 0009-0007-3760-6080

Lea Katharina Michel, Information Systems, University of Siegen, Siegen, Germany,
ORCID: 0009-0005-3559-6347

Thomas Ludwig, EduTech for Digital Transformation, University of Hagen, Hagen, Germany,
ORCID: 0000-0003-4020-6321

The restaurant industry is currently facing a challenging socio-economic situation caused by the rise of delivery services, inflation, and typically low margins. Often, technological opportunities for process optimization or customer retention are not fully utilized. In our design case study, we investigate which technologies are already being used to improve the customer experience in restaurants and explore a novel new approach to this issue. We designed, implemented, and evaluated a platform with customers and restaurateurs to increase visibility and emotional connection to nearby restaurants through their dishes. Some of our key findings include the enormous potential of combining location-based systems and conversational agents, but also the difficulties in creating content for such platforms. We contribute to the field of Human-Food Interaction by (1) identifying promising design spaces as well as customer and restaurateur requirements for technology in this domain, (2) presenting an innovative design case study to improve the user experience, and (3) exploring the broader implications of our design case study findings for approaching a real-world metaverse.

**Additional Keywords and Phrases:** Human-Food Interaction, Location-Based Services, Conversational Agents, Augmented Reality, Real-world Metaverse

## 1 INTRODUCTION

Restaurants have long faced the challenge of maintaining profitability due to their inherently low margins (Tan and Netessine, 2020). More recently, the need for further adjustments has become apparent, requiring actions such as adopting cashless payment methods, complying with safety regulations, and integrating or expanding existing food delivery infrastructure (Karniouchina et al., 2022). A significant portion of restaurants struggled to adapt to evolving social expectations and the digital landscape, and ultimately were unable to overcome these challenges (National Restaurant Association, 2020). However, businesses that successfully cultivate robust customer loyalty significantly improve their prospects for long-term. Restaurants that merely offer discounts do not experience a commensurate reduction in the risk of closure (Karniouchina et al., 2022). This underscores the importance of focusing on the long-term customer loyalty, rather than relying solely on short-term and discount-driven strategies (Karniouchina et al., 2022). As a result, it becomes compelling to look at the entire dining experience from a customer journey perspective (Lemon and Verhoef, 2016).

Within our study, we examine how restaurateurs can use innovative technology to engage with potential customers and guests and improve their experience regarding the restaurant itself. We therefore conducted a design case study in which we uncovered design spaces by adapting a participatory design approach in which we asked restaurateurs about their use of technology and media through semi-structured interviews, and involved both restaurateurs and restaurant guests in the actual design process. Through that, we gained an understanding of the diverse landscape of opportunities, challenges and innovations in restaurant experience design from an HCI perspective. Based on the derived requirements, we implemented the *Rendezfood* platform that enables new restaurant experiences by combining location-based services and proactive anthropomorphic conversational agents (CAs) with food. Following the large research discourse of human-food-interaction (HFI) as a subset of HCI, our goal was to foster technology-based emotional connections between guests and food while providing innovative benefits (e.g., additional information about the preparation of the meal or the origin of the ingredients). To examine the strengths and weaknesses of the platform, we conducted two evaluation studies with four restaurateurs and with nine restaurant guests.

By reflecting on our design case study, our paper provides three main contributions with a special focus on HFI: we explore (1) what requirements and needs restaurant guests and owners have for current technologies during dine-in situations at restaurants; (2) how an innovative approach for improving the guest (user) experience during restaurant visits might look like; and (3) what future approaches seem to be promising for the years ahead based on the lessons learned from our design case study and in respect to current developments around the real-world metaverse (Hanke, 2021; Xu, 2022).

## 2 RELATED WORK

Our research combines insights from the fields of eating out practices, specifically dining-in situations at restaurants, and current Human-Food Interaction (HFI) interventions to support eating out practices through technology. We analyze the practice of eating out through a temporal lens that focuses on a customer journey following a restaurant guest through a typical eating out experience.

### 2.1 The Practice of Eating out

Whereas in the past a visit to a restaurant was more of a celebratory nature, this changed in the mid-20th century with the increasing availability of ethnic foods and the associated greater choice, leading to a democratization of eating out for society (Warde, 2018). Thus, Paddock et al. (2017) concluded that by 2015, eating out had become a "dominant" practice in the UK, practiced informally and routinely. A detailed look at the practice of eating out is provided by Wenzer (2010), who observed and documented different eating out practices of young Swedish school students in an ethnographic field study. In addition to the practice of "replacing-school-meal-with-other-food", the social character of eating out is particularly evident in the described practice of "hanging out", where only a small snack or drink is ordered to be allowed to sit with friends in the venue. With "fika" he describes a practice in which the actual eating becomes secondary and socializing with friends or reading a book while eating is pursued as the main activity (Wenzer, 2010).

The social nature of eating out is also underscored in the study by Weber et al. (2020), who show that young people in Germany rarely eat alone and that young adults have a strong aversion to eating out alone. In addition, the study highlights and addresses the ubiquity of technology, especially smartphones, although there is also a strong reluctance to use smartphones when eating out with others. In contrast, respondents have little concern about using their smartphones when eating alone (Weber et al., 2020). This shows the extent to which smartphones can negatively affect the positive feeling of perceived commensality. While commensality is basically defined as eating at the same table (Fischler, 2011), it plays a key role in strengthening and forming social bonds in groups, especially families, through eating and sharing food together (Fischler, 2011; Ochs and Shohet, 2006). However, Koponen and Mustonen (2022) show that participants reported enjoying and appreciating food more when eating alone than when "distracted" by social interactions. The study also indicates that people are willing to reduce their use of technology when eating alone in restaurants to focus more on their food (Koponen and Mustonen, 2022).

In the past, it has also proved beneficial to examine eating out practices in more detail and to take a closer look at the temporal sequence of individual phases, which describe the use or role of technology in eating out practices (Berezina et al., 2019; Kimes, 2008; Weber et al., 2020). It should be noted that the identified phases of these studies are based on different approaches and have different objectives. As such, Berezina et al. (2019) identified chatbot capabilities, Weber et al. (2020) understood guests' use of technology, and Kimes (2008) focused on presenting the role of technology from/for a management perspective. However, there is a lot of overlap between them (Table 1).

**Table 1: Comparison of different terminologies used to describe the practice of eating out at a restaurant.**

| Authors | Terminology for Different Phases of the Restaurant Customer Experience | | | | |
|---|---|---|---|---|---|
| Berezina et al. (2019) | Pre-arrival | At the Restaurant | | | Post-dining |
| Weber et al. (2020) | - | Before Dining | | While Dining | After Dining |
| Kimes (2008) | Prearrival | Postarrival | Preprocess | In-process | Postprocess | Table turnover |
| The Terminology used in this paper | Prearrival | Postarrival & Preprocess | | While Dining | After Dining |

In the context of our paper, a combination of Weber et al.'s (2020) and Kimes' (2008) conceptual models seemed most appropriate. While Kimes' (2008) classification is the most granular, the managerial perspective brings unneeded granularity to the *After Dining* phase by separating between *Postprocess* and *Table turnover*. Thus, the combination of Weber et al.'s (2020) comprehensive model with Kimes' (2008) managerial perspective ensures a complete yet comprehensive view of the role of technology in dining-in situations. Thus, the actions and technology used up to entering the restaurant fall into the *Prearrival* phase, everything after that up to tasting the food or beverage we group into the *Postarrival & Preprocess* phase. As long as people are actively eating or drinking, we refer to it as the *While Dining* phase. The process of paying, leaving the restaurant and reflecting on the experience is considered under *After Dining*. Looking at the practice of eating out with the four subphases allows us to present and reflect on our findings in a structured way. We also use the four phases in the following chapter to introduce HFI interventions for eating out and to position them within the phases.

## 2.2 Human-Food Interaction Interventions for Eating out

Human-Food Interaction, as a subfield of HCI research, encompasses not only the study of eating practices, but also the purchase, cultivation, storage, tracking, preparation, and disposal of food (Altarriba Bertran et al., 2019; Khot and Mueller, 2019). Nevertheless, eating practices play a central role in HFI research, as reflected in their dominant position as one of the six main domains in Altarriba Bertran et al.'s (2019) framework and as one of the four main themes in Khot and Mueller's framework (2019). However, Altarriba Bertran et al. (2019) point out that much of the research in the domain of eating is concerned with individual eating experiences. There is a predominant emphasis on the perception and influence of taste, particularly driven by the multisensory HFI community, while "papers with a social approach are still less common" (Altarriba Bertran et al., 2019). Since eating out is rarely done alone (Martens, 1997) and can cause discomfort (Goode, 2018; Weber et al., 2020), the domain of "eating out" is an inherently social field of tension. Even if one enters a restaurant alone, there are usually strangers at the other tables, making the restaurant a public space that functions as a "third place" (Oldenburg, 1999). To see how this social field of tension has been addressed, the following sections provide an overview of HFI interventions along the four phases of eating out.

### 2.2.1 Prearrival

When people decide to eat out, they are faced with two major questions: "What do we eat?" and "Where do we eat?" First, digital restaurant review systems can have a significant impact on the destination. Today, consumers prefer rating systems such as Google Maps Review over analog information (Lee and Kim, 2020). Rating systems, such as star or point systems, and more detailed, written judgments in the form of reviews by previous restaurant guests can give users an impression of the quality of a restaurant before they even visit it. In systems such as Google Maps, this is linked to a rating system for each location. With the ubiquity of smartphones and location-based services (LBS), users can see what dining options are available in their area, how to get to the restaurant of their choice (Steiniger et al., 2006), make seat reservations (Schaarschmidt and Höber, 2017), or get nutrition values and allergens of restaurant food (Mandracchia et al., 2022).

In recent years, we have seen an increase in the use and trust of social recommender systems, which provide users with one or more restaurant suggestions based on their preferences (Bakhshi et al., 2014). This also means that some apps (e.g., *HappyCow* (HappyCow, 2023) or *Find Me Gluten Free* (FindMeGlutenFree, 2023)) consider food intolerances, allergies and special diets (e.g., gluten-free or vegan) when recommending restaurants (Mandracchia et al., 2020). There are even promising approaches that combine conversational agents (CAs) with such recommender systems, resulting in a conversational recommender system (CRS) (Sardella et al., 2019). For example, while the Telegram chatbot developed by Sardella et al. (2019) uses a collaborative filtering approach and recommend restaurants based on the user's location and time of day, AllergyBot (Hsu et al., 2017) first asks about the user's allergies and then recommend suitable restaurants to the user that do not conflict with the allergies. When eating out as a group, the *what* and *where* often need to be negotiated. To support this negotiation process, Terzimehić et al. (2018) developed a Skype chatbot to guide groups through the process of deciding where and what to have for lunch. The chatbot simplifies the decision-making process and reduces social pressure by anonymizing votes and making users' votes count more in the next election if they previously decided against their own choice (Terzimehić et al., 2018).

### 2.2.2 Postarrival & Preprocess

Once (a group of) customers arrive at the restaurant, there are several ways that technology can support the *Postarrival & Preprocess* phase of the eating out practice. Most attempts have been to digitalize and optimize the ordering process, but also to enrich it with additional information compared to a conventional menu card. While the use of tablets already makes it possible to order food digitally via a digital menu or to see which dishes or beverages are currently sold out (Lessel et al., 2012), enabling restaurants to reduce ordering times and costs (Susskind and Curry, 2019; Tan and Netessine, 2020), there are approaches that build on this and aim to create special experiences through the use of augmented reality (AR): Either through the use of Mobile Augmented Reality (MAR), where the smartphone is used as a lens that can be held over the menu (Arioputra and Lin, 2015; Lee et al., 2020) or table (Garcia, 2023; Lee et al., 2020) to allow for more interaction, or through projections created by projectors mounted on the ceiling (Margetis et al., 2013; Shengzhi, 2015; Sterckx and Verbeeck, 2023). In this way, it is possible to preview and order dishes (Lee et al., 2020), translate the menu (Arioputra and Lin, 2015), and shorten the waiting time with small humorous animations (Batat, 2021; Sterckx and Verbeeck, 2023) or breakout type games using empty plates (Margetis et al., 2013). Even the choice of food can be influenced by such systems, for example to eat more local food (Lee et al., 2020). Robots can be used to interact with customers in the restaurant, such as taking and serving orders (Asif et al., 2015; Garcia-Haro et al., 2020) or providing entertainment (Pieska et al., 2013). To make restaurant visits more accessible, Obiorah et al. (2021) designed and evaluated three prototypes to allow people with aphasia to explore menu items and their ingredients, which is otherwise difficult for this user group.

In addition to the opportunities, however, it is also important to highlight the risks associated with the use of technology in this phase, which have not yet been fully explored. In any case, with regard to the use of digital menus, the initial expected negative outcomes in the study by Lessel et al. (2012) were a decrease in social interaction with service staff and a deterioration of the atmosphere in the restaurant, since it is "a place that should not be digitized" (Lessel et al., 2012) per se. Technology adoption of robotic systems in this context can be improved mainly by increasing trust and lowering the perceived risk of use (Choe et al., 2022; Seo and Lee, 2021).

### 2.2.3 While Dining

*While Dining* out, technology can potentially be used to create new and enhanced culinary experiences (Gayler, 2017; Heller, 2021; Spence, 2020; Wang et al., 2019), cope with eating alone (Jarusriboonchai et al., 2014; Khot et al., 2019; Weber et al., 2020; Wei et al., 2011), encourage sustainable behavior (Genç et al., 2019), assist service staff (Rocha et al., 2011), and make this phase more accessible for people with disabilities (Chung et al., 2021). In terms of new and enhanced dining experiences, Spence (2020) describes how some restaurants and cafes are pairing music and soundscapes via headphones and earbuds with the theme of the food to enhance the overall experience. Sounds also play an important role in Wang et al.'s (2019) design, as they are played in *iScream!* when the ice cream is licked, creating a "novel gustatory experience" (Wang et al., 2019). By focusing on different colors of light coming from the table, Burkert et al. (2022) show that it is possible to manipulate taste perception with their *ColorTable* prototype. Somewhat differently, the use of ephemeral edible interfaces, such as 3D-printed food (Gayler, 2017) or conductive materials like pastry (Heller, 2021), have the potential to create new types of interactive experiences.

Alternatively, technology can potentially help to cope with eating alone. For instance, a sense of digital commensality (Spence et al., 2019) can arise by digitally eating with other people (Wei et al., 2011), interacting with robots (Khot et al., 2019; Mancini et al., 2020) or virtual agents (Weber et al., 2021). But there are also approaches, as described in (Jarusriboonchai et al., 2014), that bring like-minded people together on the spot, for example by allowing positive thoughts or pictures from the café to be shared with nearby guests.

Other approaches, such as the *adaptive buffet* (Genç et al., 2019), aim to reduce the number of leftovers at buffets and thus contribute to sustainability. A smart coaster, on the other hand, assists service staff by informing them when the drinks ordered are empty, so they can immediately ask if another drink should be ordered (Rocha et al., 2011). In addition, there are other initiatives to help people with visual impairments by providing them with different audio cues to communicate where exactly the food is on the table (Chung et al., 2021).

*2.2.4 After Dining*

In addition to the integration of various contactless payment methods (Liyanage et al., 2018; Patil et al., 2017), which can optimize guest satisfaction by minimizing waiting times and facilitating the management of sales data (Susskind and Curry, 2019; Weber et al., 2023), we found little recent relevant interventions in the *After Dining* phase. Since customers often share pictures or write restaurant reviews during this phase (Weber et al., 2020), there is a preliminary approach to using facial recognition to automatically compose restaurant reviews (Mahrab et al., 2021). This could be beneficial for recommender systems and for verifying the authenticity of reviews, but it also poses the same privacy and data protection risks as other applications of facial recognition in public spaces (Fontes and Perrone, 2021; Zalnieriute, 2021). In this phase, users can also be provided with a report that converts the food consumed into the miles it took to transport it, to encourage more sustainable eating habits in the future by consuming more local products (Lee et al., 2020).

## 3 RESEARCH QUESTION AND APPROACH

Overall, the research around HFI interventions in the multifaceted practice of eating out has increased in recent years, but still offers much room for further investigation. The segmentation of the eating out practice into the four phases *Prearrival, Postarrival & Preprocess, While Dining* and *After Dining* seems to provide the necessary structure to better understand and classify existing interventions in terms of their impact on the user experience and to design future experiences with this in mind. We agree with Covaci et al.'s (2023) assessment of general HFI research that "[…] existing HFI solutions work in isolation, […] which could limit user's agency, the focus of the experience, and the domain of intervention.", as we noted that current interventions for eating out often focus on isolated phases of the restaurant visit. There are only a few interventions that address more than one phase of the eating out practice (e.g., (Batat, 2021; Lee et al., 2020)). Additionally, the last two phases (*While Dining* and *After Dining*) seem to receive less research attention. While the quantity of interventions listed in the *While Dining* phase appears promising, it should be noted that some of the interventions (especially early prototypes) (Heller, 2021; Mancini et al., 2020; Wang et al., 2019; Wei et al., 2011) were not (yet) explicitly designed or in-situ evaluated for the eating out context in their original studies and were investigated in laboratory settings, but seem to have a high potential for application during in this phase.

Across all phases, it is striking that most research focuses on conceptualization, design, and technical development, with limited emphasis on evaluations to validate the approaches developed. Especially, evaluations in real-world contexts are lacking, with a few exceptions (e.g., (Batat, 2021; Obiorah et al., 2021; Terzimehić et al., 2018)). Thus, we address three research questions with our work:

- **RQ 1:** What are the needs and requirements of restaurant guests, restaurateurs and their employees for current technologies in the restaurant environment?
- **RQ 2:** How can an innovative approach be designed that addresses the needs in and for a real-world settings?
- **RQ 3:** What design approaches might be promising for the coming years based on the lessons learned from our design case study?

We answered the three RQs by conducting a three-phase design case study (Wulf et al., 2011), where we first explored the social practices of restaurant customers and restaurant owners in relation to their technology use. Through subsequent co-design sessions we created user experience stories with restaurant owners and evaluated the stories with restaurant customers. Based on the evaluations, we uncovered promising design opportunities and on the basis of these opportunities, we developed an ICT artifact guided by the "provotyping" approach (Boer and Donovan, 2012). We adopted the interaction concept of "edible anthropomorphic virtual agents" (EAVAs) (Weber et al., 2021), made them proactively triggered in restaurants in a location-based manner, and allowed for customization by the restaurant owners. The evaluation of the developed system highlighted the provocative qualities of this intervention, which resulted in rich insights and a deeper understanding of how future technological interventions can be designed to be socially acceptable. It also highlights promising approaches to enriching the guest experience that are worth pursuing.

## 4 USER STUDIES

We first conducted an empirical study to gain a better understanding of current social practices and, in particular, the use of technology in dining-in situations at restaurants. Our study focuses on how restaurant owners use technology and media to attract and retain customers. With our study, we supplement the findings of Weber et al. (2020) who explicitly examined how technology is used by restaurant guests. In subsequent co-design sessions, we worked with the restaurant owners to elaborate on the interview results, with a focus on identifying design opportunities through user experience (UX) stories, and then piloted the developed UX stories for acceptance by restaurant customers.

### 4.1 Methodology and Participants of the Interview Study

To understand how technology and digital media are used in gastronomy to attract and retain customers, we conducted seven semi-structured interviews with restaurant owners in 2019. We interviewed local restaurateurs, making sure to include a variety of perspectives. This gave us a wide range of experiences, from the restaurant owner who runs a restaurant specializing in innovative French fries, to the owner of a café with three locations, to the owner of a supermarket chain whose stores offer warm lunches for on-site consumption (Table 2). We conducted all interviews on-site at the interviewees' venues. In the interviews, we employed a combination of predetermined and open-ended questions to facilitate further inquiries, following the recommendation of Gall et al. (2003). We encouraged an open discussion about eliciting good and bad experiences with technology and media to better understand them. The interviews were then fully transcribed and analyzed by the author team using the hybrid approach of Fereday and Muir-Cochrane (2006) to apply deductive codes based on the interview guide, which subsequently resulted in inductive codes from the collected data. The coding framework was reviewed and modified in multiple sessions among the authors. Within this paper, we translated the German quotes into English as close to the original as possible.

Table 2: Restaurateurs interviewed and their role in the business.

| Participant | Gender | Age | Role in the company |
|---|---|---|---|
| R1 | M | 24 | Co-owner of a frozen yogurt and snack bar |
| R2 | M | 43 | Owner of three café and ice cream parlors and an online shop |
| R3 | M | 30 | Co-owner of two restaurant specializing in innovative French fries |
| R4 | M | 44 | Owner of a regional supermarket chain, including bakeries and simple lunches for on-site consumption |
| R5 | M | 28 | Co-owner of a café and ice cream parlor |
| R6 | M | 32 | Owner of a restaurant specializing in venison/game dishes |
| R7 | F | 45 | Owner of a café |

### 4.2 Results of the Interview Study

In this section, we explore the current landscape of technology used by restaurateurs. Several themes emerged from the interviews, primarily related to the *Prearrival* phase and digital advertising challenges. Below is an overview of the existing technology usage and its nuances among restaurant owners.

Digital marketing, particularly through social media, is a cornerstone of restaurant owners' advertising strategies. The most widely used platforms in this area are Facebook and Instagram. These channels are preferred over traditional newspaper advertising because of their cost-effectiveness and ability to reach highly targeted audiences. One striking observation was the shift toward visual media, with the realization that images have a unique ability to evoke an emotional response and generate interest in food. As one restaurant owner put it, "A picture like that touches you in a much more emotional way than when you read something like '*Fresh fries today!*'" (R3). In addition, all of the restaurant owners interviewed expressed the importance of posting regularly on social media. They adhere to posting schedules and occasionally respond to external events or factors with impromptu posts or stories, emphasizing the need for consistent online engagement.

Despite their reliance on social media platforms, restaurant owners expressed doubts about the effectiveness of paid social media advertising. One even noted a decline in organic reach and suspects the presence of algorithms designed to drive ad spending. He explained, *"As soon as I advertise, it [the number of* *likes] goes up. But I think there is simply an algorithm that is supposed to pull our money out of our pockets"* (R2). In addition, restaurant owners questioned the relevance of having followers from distant locations, when they operate primarily as local businesses and emphasized the importance of reaching nearby regional audiences (R2).

Customer engagement is an important aspect of restaurant marketing. Restaurant owners actively encourage customers to share their experiences and opinions through comments, reviews, and feedback on social media and review platforms such as Google. These reviews carry significant weight and directly influence the evaluation of marketing campaigns. While discounts and promotions are rarely a core part of the marketing strategy for several restaurant owners (R1, R2, R6,) and have not been very successful in the past (R6, R7), R5 has very much integrated discounts and promotions into his strategy. However, he wants to move towards automated systems for managing these promotions, recognizing the need for more streamlined and efficient processes.

Maintaining a digital presence is considered vital, and restaurant owners frequently update their websites to provide information on weekly specials and career opportunities (R4). While not actively pursued by all, search engine optimization (SEO) and search engine advertising (SEA) are areas of interest and investment for future improvement (R4, R5, R6).

However, restaurants also emphasized the importance of offline advertising methods, as they use various offline advertising methods to reach a wider audience. Traditional paper flyers distributed with newspapers play an important role (R2, R3, R4, R6, R7). In addition, radio advertising is used periodically during peak seasons (R2, R4). R3 was notable for its active participation in street food festivals and city events, demonstrating a proactive engagement with the local community beyond traditional restaurant settings. There is also a strong tendency towards creative offline advertising solutions, such as placing signs on trees (R6) or distributing pieces of artificial grass with their branding (R1) and handing out appetizers to passers-by (R1). Noteworthy, R5 shared a unique approach of deliberately limiting online menu information to drive physical visits, consistent with a deliberate strategy of creating a sense of wonder around the dining experience.

Restaurant owners (R1, R2, R5, R6, R7) also acknowledged their strategic adaptation of offerings and marketing strategies based on weather conditions, demonstrating a keen awareness of environmental factors and seasons that influence customer behavior. Interestingly, R3 and R4 expressed curiosity about experimenting with such adaptive approaches, indicating a willingness to explore innovative practices. It is also interesting that none of the restaurant owners was working with influencers. R3 expressed a principled reluctance, citing concerns about potential inauthenticity, while R5 and R7 remained open to partnerships as long as the influencers maintain authenticity and credibility in their engagement.

Moving to the *Postarrival & Preprocess* phase, a common thread emerged in the emphasis on personalized communication by restaurant owners R1 and R5. The practice of using customers' first names was highlighted as a means of fostering a friendly and welcoming atmosphere within the establishment. Interestingly, at the ordering stage, the majority of restaurant owners maintained a traditional approach, using software primarily for order processing and billing, but not at the customer touchpoint. Electronic payment methods had not been widely adopted at the time of the interviews, reflecting a prevalent reliance on traditional methods within this cohort. In particular, there was a perceived resistance to the integration of technology into customer interactions, particularly highlighted by R3. The sentiment revolved around a concern that technology might interfere with the interpersonal connection between customers and service staff. An interesting observation was made several months after the interview with R3 during the Covid-19 pandemic, when his restaurant introduced a terminal for ordering meals. Although the reasons for introducing a terminal were not covered through subsequent interviews, it underscores the industry's adaptability to external challenges, especially given the accelerated adoption of electronic payment methods by restaurants and customers during the pandemic (Weber et al., 2023). The interview study highlighted a predominant focus on technology, particularly in the *Prearrival phase* and (to some extent) in the *Postarrival & Preprocess* phase, with a strong focus on marketing efforts.

### 4.3 Co-creation and Evaluation of User Experience Stories

To visualize initial design ideas and to directly iteratively refine or discard them, we derived 18 small UX stories (Quesenbery and Brooks, 2010) through co-design sessions with restaurateurs and then evaluated them in a UX speed dating workshop (Davidoff et al., 2007; Zimmerman and Forlizzi, 2017) with potential users. In the co-design sessions, five restaurateurs from the first user study were able to join us in considering what features and functionality a future platform should have. This included considering what role emerging technologies such as LBS, CAs or AR could play. Participants also had the opportunity to visualize their ideas. We refined the visualizations and visualized and communicated the core concept of each UX story in the UX speed dating workshop (Figure 1). We also included elements of innovative current research in the UX stories to discuss with the participants. For example, we explained the possibility of filtering and searching for the right dish depending on one's allergies (derived from the AllergyBot concept (Hsu et al., 2017)), and also discussed the concept of EAVAs (Weber et al., 2021) as a possible way of interacting with dishes, explaining them as food avatars and "chatbot-empowered meals" in a way that was easy to understand. Eight participants (aged 16 to 22) took part in the UX speed dating workshop and had the opportunity to comment on and rate individual user experience stories. These two steps helped us to generate and develop ideas with restaurant owners as well as restaurant guests and to assess which functionalities should be prioritized.

After presenting the created UX stories to the eight participants, it became clear that they wanted an easy way to find suitable dining options in their area, ideally within their budget. In addition, the current waiting time for restaurants should be displayed. To prevent the location-based approach from leading to too many notifications, it should also be possible to manually deactivate the notification of unwanted restaurants or food chains right from the start, and a temporary deactivation of all notifications is also desired.

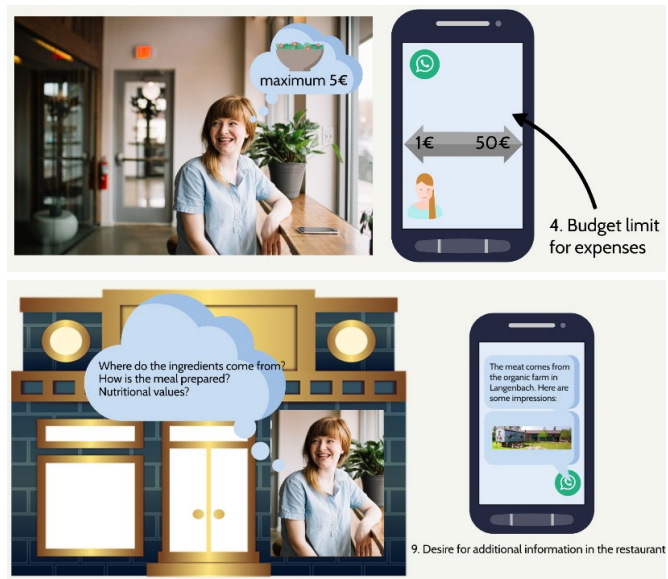

Figure 1: Visualization of UX stories 4 (budget-based filter) and 9 (additional information in restaurants through a chatbot)

### 4.4   Identifying Design Opportunities

Our interactions with restaurant owners revealed that technology plays a critical role in attracting and retaining customers. Social media platforms such as Facebook and Instagram are widely used for marketing, primarily due to their low cost and wide audience reach. However, there is skepticism about the effectiveness of paid social media advertising, with concerns about manipulated views and clicks. Restaurant owners expressed the need to target local, regional audiences and emphasized the importance of visual media in attracting potential customers.

Based on the findings of our interview study and the co-design sessions with restaurateurs, as well as the feedback we received from the evaluation of our UX stories through UX speed dating, we derived several design implications for future interventions in the restaurant domain. First, it is critical to emphasize visually appealing and engaging content to capture users' attention, which was repeatedly mentioned in different contexts during the evaluation of the UX stories and in the interviews of the restaurateurs. Second, interventions should incorporate geolocation-based targeting mechanisms to address restaurant owners' concerns about reaching a nearby regional audience. By prioritizing nearby users and providing relevant recommendations, interventions can effectively target the desired audience. Third, improving discoverability is essential for restaurants. Restaurants should use eye-catching advertising and optimize their physical space to attract potential customers. Digital interventions can support this by incorporating interactive elements such as proactive CAs. Fourth, future interventions should not be perceived as a mere marketing gimmick, but also offer added value such as personalization and customization options as users have different preferences, dietary restrictions, and budget limitations. By allowing users to customize their profiles and receive tailored recommendations, interventions can provide personalized experiences. By addressing these considerations, we aimed to create a user-centered and engaging eating out experience that meets the needs and expectations of both restaurant owners and customers.

## 5 DESIGN OF ICT ARTIFACTS

### 5.1 Interactive Prototype

Based on the identified design opportunities of the explored design space, we developed a coherent interactive prototype that combined different concepts from the UX stories and feedback from the UX speed dating workshop, and served as the basis for the development of the functional app called *Rendezfood*. Next, we provide an overview of the basic features of the app.

In the *Prearrival* phase, a user's profile can be set up to specify allergies, intolerances, but also preferences for ingredients or diets (vegan, vegetarian, meat), as well as a budget limit not to spend more than a certain amount of euros. Restaurant chains and specific restaurants can be blacklisted to avoid receiving notifications from these restaurants (Figure 2a). Now, when the user is near a restaurant, chatbot-empowered meals from the restaurant can proactively notify the user based on the user's location (Figure 2b). Meals can also be freely explored via a vertically scrolling feed (*Explore screen*) (Figure 2c) that displays each meal with an image and its ingredients, which serves as the home screen when a user launches the *Rendezfood* app. Depending on the respective profile settings, three matching meals from restaurants nearby are always recommended in a sperate screen (*Exploit screen*) (Figure 2d). By designing the two screens differently, we deliberately want to give users the option of comfortably scrolling through the meals using the *Explore screen* if they have time to do so. On the other hand, if the user has little time to decide, the *Exploit screen* should allow for a quick selection. The terms *Explore* and *Exploit screen* are derived from the concept of the "explore/exploit tradeoff", which describes "the balance between trying new things and enjoying our favorites" (Christian and Griffiths, 2016). It is also possible to find and navigate (via an external app) to the restaurant associated with the meal.

In the *Postarrival & Preprocess* phase, it is still possible to talk to the chatbots to learn more about the ingredients, the origin of the food or the nutritional values of the meal, as well as to interact with the meal in a playful way, e.g., by asking for a joke or triggering other fun interactions (Figure 2e). In the *While Dining* phase, a QR code can be placed in the meal. When the user activates the AR feature, virtual eyes, arms and a mouth are placed on the corresponding parts of the real food according to the placed QR code (Figure 2f). Responses to touch interactions with the avatar, such as touching the food's mouth, eyes, or hands, will then be supported, and the food avatar will read aloud the text output it provides. In the *After Dining* phase, it is still possible to converse with the chatbot of the meal eaten or with other meals.

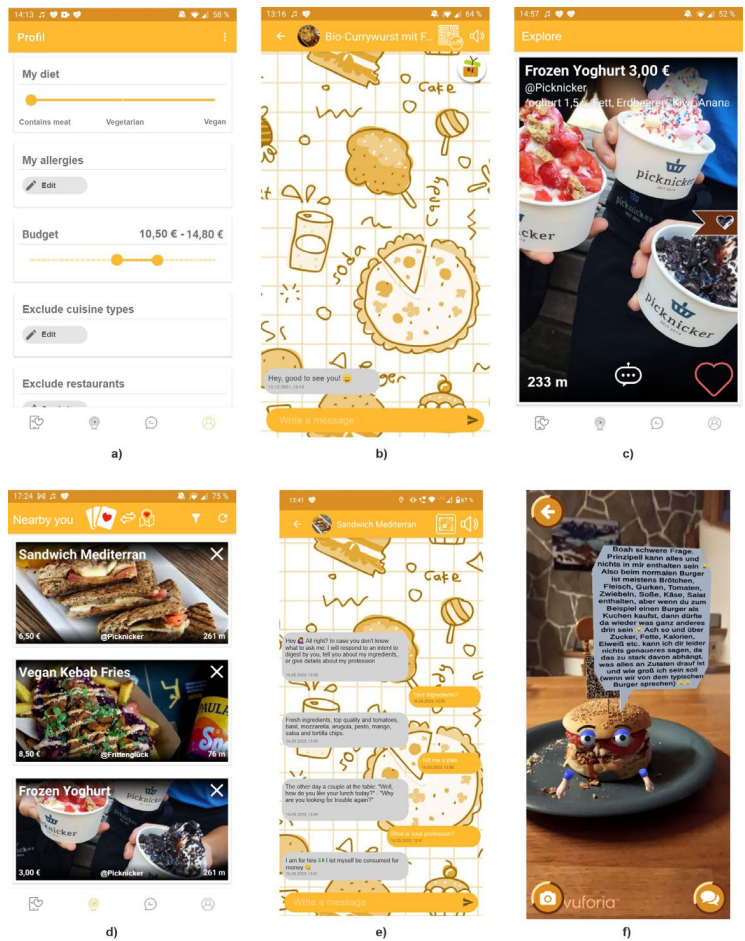

Figure 2: Overview of the interactive Click Prototype Features

## 5.2 Mobile App for Restaurant Guests

We implemented the *Rendezfood* app as a native Android application that aimed to enhance the dining experience by incorporating location-based notifications, intent-based chatbots powered by Dialogflow, a middleware for dynamic data, an embedded Unity application for AR capabilities, and a recommendation engine for meal suggestions. To enable location-based notifications from the chatbot, the app initially required users to grant permission for location access. This feature ensured that users received relevant notifications based on their proximity to restaurants, allowing for a more personalized dining experience.

While free text input was allowed, quick response buttons were not provided. This approach allowed for more flexibility in user interactions and accommodated a wider range of user questions and requests. We implemented a middleware component to enhance the chatbot's responses with dynamic data. This included incorporating current time information or providing fun answers based on special food days or meal nicknames.

The mobile app included an embedded Unity application that seamlessly integrated with the chatbot interface, enabling AR interactions with the food. The AR feature is enabled through QR code tracking. When opened by clicking a button, a virtual food avatar provided voice output with a male, female, or random (if gender was not specified) voice based on the gender selected by the restaurant owner for that meal. The mobile app incorporated a recommendation engine to provide meal suggestions on the exploit screen. Based on the user's profile settings, such as allergies, dietary preferences, and budget limits, the app recommended three matching meals from nearby restaurants.

### 5.3 Web Interface for Restaurateurs

We developed a web interface specifically designed for restaurateurs to facilitate the management of their restaurants and menus. The web interface consists of several microservices that provide the frontend developed with Angular (HTML5, SCSS and JavaScript), the backend developed with Java and C# / ASP.NET Core, and the database. The main focus this interface is to provide a user-friendly and intuitive authoring platform and high-level content design framework. One of the key features of the platform is the ability to effectively manage restaurant locations. Restaurateurs can easily add, edit, disable, or delete restaurant locations as needed. They also have the flexibility to adjust the hours of operation for each restaurant on different days. The interface also allows restaurateurs to define geofences, enabling them to trigger automatic push notifications to users who are in proximity to their restaurant.

Another key aspect of the web interface is the comprehensive dish management system (Figure 3). Restaurateurs can easily add, edit, temporarily disable, or remove items from their menus. When adding a new dish, restaurateurs have the option to provide various details, including the dish's name, nickname, image, ingredients, description, price, gender specification (male, female, or unspecified), allergen information, cuisine type, and the use of local and/or organic ingredients. Restaurateurs can also indicate whether the dish is vegan, vegetarian, or contains meat. In addition, the interface allows restaurateurs to assign seasonal swipe effects to each dish, such as Spring, Easter, Summer, Fall, or Winter, which are applied when a user swipes on their dish on the *Explore screen*. In addition, restaurateurs can create separate geofences for specific dishes or use the default geofence associated with their restaurant. In addition, the web interface offers integration with a Unity web plugin, which was initially developed in (Krings et al., 2022) and that allows restaurateurs to customize the virtual food avatar associated with each dish.

In addition, the web interface includes a monitoring system that allows restaurateurs to easily track key performance indicators (KPIs), including metrics such as most talked to dishes, most popular items, trending local menu items, and user statistics that include registered and active users.

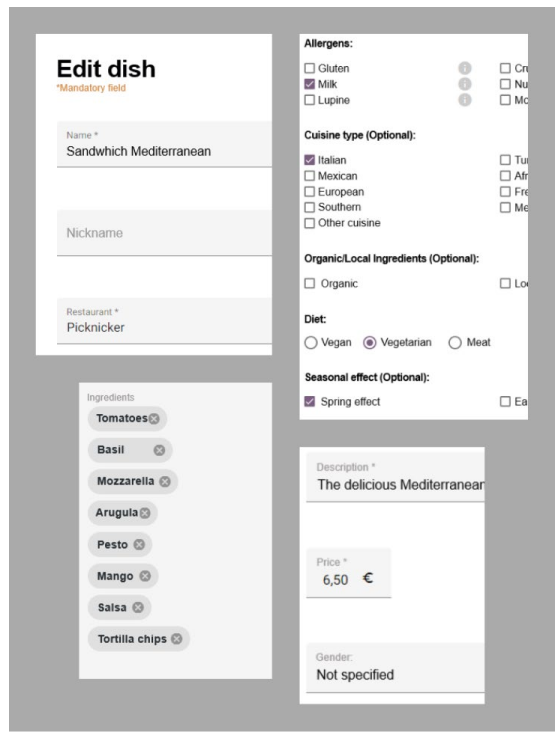

Figure 3: Highlighted parts of the web interface for editing a dish

## 6 EVALUATION

We conducted an exploratory real-world evaluation of the Rendezfood mobile app with nine participants. Sessions included individual and paired evaluations, with observations, semi-structured interviews, and app data collection. We also evaluated the web interface in four sessions with restaurant owners. The sessions evaluated content creation features, relying on observations, the think aloud approach, and post-session interviews.

### 6.1 Evaluation I: Mobile App Setup and Applied Methods

We conducted the evaluation from the end of September to the end of October 2021. Sessions lasted between 1 hour and 40 minutes and 2 hours and 5 minutes. Weather conditions during the study period were mostly dry, with occasional light drizzle and temperatures ranging from 11 to 20 degrees Celsius. Nine participants (*Guests*; G01-G09) were recruited through announcements at our faculty, word of mouth, and snowball sampling. As an incentive, a free meal was offered as part of the study. To observe potential group dynamics, the study included both individual sessions (G01, G02, G03) and paired sessions (G04 & G05, G06 & G07, G08 & G09) (Table 3).

Table 3: Basic data of the participants and the evaluation sessions

| Guests | Gender | Age | Dietary restrictions | Weather conditions |
|---|---|---|---|---|
| G01 | M | 29 | No restrictions | Dry, 20 °C / 68 °F |
| G02 | W | 24 | No restrictions | Dry, 15 °C / 59 °F |
| G03 | M | 31 | Fructose intolerance | Drizzle, 15°C / 59 °F |
| G04 | W | 23 | Vegetarian | Drizzle, 12°C / 53.6 °F |
| G05 | W | 24 | Vegetarian | Drizzle, 12°C / 53.6 °F |
| G06 | W | 26 | No restrictions | Dry, 11°C / 51.8 °F |
| G07 | W | 23 | No restrictions | Dry, 11°C / 51.8 °F |
| G08 | M | 27 | No restrictions | Sunny, 13°C / 55.4 °F |
| G09 | M | 23 | No restrictions | Sunny, 13°C / 55.4 °F |

Two researchers were present in all sessions, except for one session (G03) where only one researcher was present. After a short briefing about the procedure and the tasks, the observers had a monitoring role and did not actively intervene in the participants' use of the app. They took unstructured observational notes on participants' engagement, interaction, decision making, and interactions. Participants were given the freedom to express their thoughts freely following a thinking aloud approach (Nielsen, 1994). A semi-structured interview was conducted after the dining experience to gain additional insights. In addition, video was captured from the perspective of one of the participants using a small body-mounted camera (Insta360), covering the *Prearrival* phase until arrival at the restaurant, to avoid invading the privacy of other guests. Screen recordings with audio were also captured from one device per group throughout the full session, as well as chat logs of all chatbots interacted with.

All major observations and findings were discussed directly between the researchers after the session. The handwritten observation notes were digitized and the screen recordings were synchronized with the video recordings. Afterwards, the material was carefully examined to ensure that no details were missed during the observations, and any unclear notes were clarified to guarantee an accurate interpretation. The data analysis process involved an iterative coding approach and regular discussions among the researchers to ensure a comprehensive understanding of the gathered data.

### 6.2 Results of the Evaluation Study (Mobile App)

We structure our findings along the introduced four phases of (1) *Prearrival*, (2) *Postarrival & Preprocess*, (3) *While Dining* and (4) *After Dining*. Our findings combine our observations of the participants while they were using the app and what they said in interviews after using it, as well as the recorded app data.

*6.2.1 Prearrival*

After participants launched the app, all participants decided against registering and opted for a guest account (G01-G09), as G01 explained: *"I hate when there is a need to register [...] so I would always go without registering first"* and G03 commented: *"I always like to go without registering [...]"* (G03). In addition, G01 and G09 questioned the benefits of having a profile. G01 would like to know how and by whom his personal data from his profile would be used.

Starting with the home screen, all nine participants commented positively on the *Explore screen*. Specifically, G03 found it preferable to start with the *Explore screen* to avoid being overwhelmed by having to adjust profile settings first. The scrolling interaction and visual aesthetics of the *Explore screen* were praised by G02, G08, and G09 as intuitive and user-friendly. G03 described the functionality as exciting and fun. On the other hand, G04, G08, and G09 would have preferred to be directed to the profile screen first to adjust their preferences such as budget, dietary restrictions, and food allergies.

In our study, all individuals used both the *Explore* and *Exploit screens* to make a food choice. However, the choice of restaurant and food was based on individual aspects. G01, G04, G05, and G08 chose their food because they liked the picture of it in the Rendezfood app. These four participants and G09 chose the meal before they chose the restaurant. G02 and G03, as well as G06 and G07, chose a restaurant first and then a meal, knowing the restaurant beforehand even though G02 had not eaten there before. The chatbots were used to help people choose restaurants and meals by answering questions about the price of the meal, its ingredients, or where to buy it.

Five participants (G03, G04, G07, G08, G09) questioned the differences between the *Explore* and *Exploit screens* and wanted them to be combined into one screen. This screen should include all the features of the *Explore screen* with the filter options, distance to the restaurant, and insight information about the restaurant/meal from the *Exploit screen*. G02, G03, and G07 would like to see a list of favorite meals that can be added by swiping horizontally on the *Explore screen*. In addition, G02 would like to be able to search for special meal offerings and G05 would like to see the restaurant logo as well as the name of the restaurant for a more immediate association with the restaurant.

After selecting a restaurant to go to, navigation to the restaurant was approached differently. G03, G06, G07, G08, and G09 were among those who did not use any form of navigation to the restaurant, as G03 explained: *"I know the way, what should I have done?"* (G03). The other users (G01, G02, G04, G05) in our study were looking for a map or some other form of navigation to the chosen restaurant. Each of these four individuals found the restaurant's location by tapping on a link in the app that took them to Google Maps, which they used to navigate. G01 and G08 would have liked an easier way to find the Google Maps option, and G09 was hoping for AR-based navigation.

Initially, we planned to prepare QR code sticks and have them delivered with the ordered meals at the restaurant, allowing users to interact with their food through the AR feature. However, due to the significant organizational challenges associated with this approach, especially for the various restaurant operators involved in this exploratory real-world evaluation study, we decided to hand out the prepared QR code sticks to participants at the beginning of the study. We hoped that participants would place the QR code into their food themselves, although no specific instructions were given to evaluate the potential challenges and opportunities associated with this approach. This approach led to confusion among some participants, including G01, G04, G05, and G08, who attempted to use the QR code with their smartphone camera as soon as it was handed to them. In addition, carrying the QR code was found to be impractical for G01 and G02, with G02 stating, "If I had to carry it, I wouldn't use it". These findings highlighted the need for a more streamlined and user-friendly implementation of the QR code interaction to ensure a seamless and convenient experience for users.

In general, participants enjoyed using the app during this phase, as G01 emphasized the visually appealing and vivid images of food as the app's best feature. This aspect allowed him to discover restaurants and meals he would not have considered otherwise, which he described as a *"playful approach"* to finding a restaurant. Similarly, G04, G05 and G07 appreciated the app's focus on meals and their visual presentation, as well as the filtering options to tailor recommendations to their own needs and temporary desires. Three people (G01, G03, G08) found it inspiring and practical to have food displayed nearby and to get recommendations at the meal level rather than the restaurant level. For G08, the app was particularly useful for decision making. However, participant G09 expressed that he did not personally need the kind of support provided by the app.

*6.2.2 Postarrival & Preprocess*

Upon arrival at a restaurant (and ordering at a counter, depending on the location), seven participants took seats in the restaurant. Due to a lack of seating at the restaurant, two participants (G04 and G05) chose to sit in a room at the nearby university instead. While waiting for their meal, participants engaged with the chatbot primarily for entertainment purposes. They greeted the chatbot warmly, praised its skills, and even inquired about its gender and hobbies. In addition, participants used the chatbot for informational purposes, such as for asking questions about ingredients, preparation methods, and prices. Some participants took the opportunity while waiting to explore the AR feature and the ability to chat via AR using the QR code. In particular, G02 showed enthusiasm for this functionality, engaging in an extended conversation with the chatbot via AR for over ten minutes as she eagerly awaited her meal (Figure 4, left).

Once the meal was served, many participants eagerly placed the QR code in their meal and began exploring the AR feature instead of starting their meal (Figure 4, right). As a result, some participants found the interaction distracting. Specifically, G01 and G06 found the interaction to be a distraction that delayed their meal consumption. While AR interaction at the table was found to be enjoyable, some participants expressed a desire to interact with a virtual meal while waiting for their actual meal (making it part of the *Prearrival* experience) (G06, G07) or after their meal (making it part of the *After-Dining* experience) (G07). Among the participants, G02, G04, and G09 used the AR feature extensively, while the remaining users either preferred traditional chat or had not experienced the AR feature. In addition, four users (G04, G06, G08, G09) expressed concerns about the ergonomics of holding the smartphone too far away from the served meal for an optimal AR experience. This resulted in an uncomfortable hand position while interacting with the AR chatbot. Moreover, G05 and G07 questioned the hygienic aspect of inserting the bottom of the QR code stick into their food. Two other participants (G02, G08) expressed dissatisfaction with the written text in the AR interaction, noting that the chatbot's responses disappeared too quickly (G02, G08), while their own text input remained visible for quite some time (G02).

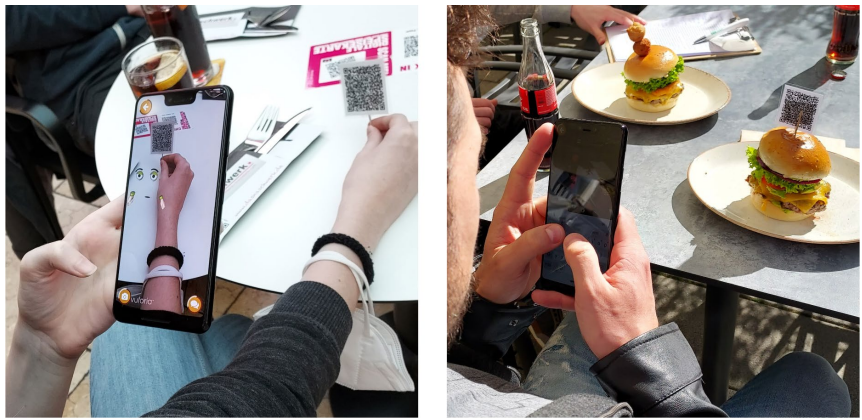

Figure 4: G02 interacting with the AR feature while waiting for her meal (left) and G01 interacting with the AR feature before starting to eat (right).

### 6.2.3 While Dining

At the beginning of the meal, three participants (G02, G04, G05) exchanged "enjoy your meal" greetings. During the meal, six participants (G02, G03, G04, G05, G08, G09) engaged in conversations with both their evaluation peer and the evaluators, discussing various aspects of the mobile app and the evaluation itself. These discussions included topics such as the features of the app, their experiences using it, and their overall impressions. In addition to app-related conversations, participants also expressed interest in each other's meals, including their taste, presentation, and overall satisfaction. This led to food-related conversations and allowed participants to share their dining experiences and preferences. In addition to these conversations, participants engaged in casual conversation, which fostered a social atmosphere during the meal. In addition, two participants, G04 and G05, delved into topics beyond the immediate scope of the study. They discussed the upcoming winter semester at the university and shared their personal life situations, moving the conversation beyond the boundaries of the mobile app and the dining experience.

While interaction with the app itself during the meal was minimal, two participants, G01 and G04, stood out for their engagement with the chatbot. G04 praised the chatbot, specifically mentioning its good taste, while G01 interacted with the chatbot to explore its reactions and interactions, especially in comparison to before he started eating it. However, such an implied response of being eaten was only very rudimentarily implemented by the chatbots in our current implementation.

### 6.2.4 After Dining

Compared to the previous phase, *Rendezfood* app usage increased again, but was still lower compared to the *Prearrival* and *Postarrival & Preprocess* phases. Chatbots were again primarily approached by users for entertainment reasons, such as asking for the name, mood, and age of the chatbot. Chatbots of other meals are also approached by asking about allergies and where the meal is located.

### 6.2.5 Comparing Chatbot Usage across the Four Phases

We analyze the chatbot usage across the four phases: *Prearrival*, *Postarrival & Preprocessing*, *During Dining*, and *After Dining*. Initially, there were 145 user inquiries generated by the nine participants, with 142 receiving a direct response from the chatbot. Three times, the bot failed to respond directly due to the user's internet connection or the user following up too quickly. Additionally, in terms of intent matching, the fallback intent was triggered 30 times, accounting for 20.7% of responses. In another 24 cases, a different intent was incorrectly matched, and in 17 other cases, the selected intent or response was only "moderately appropriate", all of which were discussed among three researchers and also interpreted in the context of the full chat history (an example of a "moderately appropriate" response is shown in Figure 5.

```
[User asked for ingredients]

Chatbot:Fresh ingredients, top quality and French fries,
beyond meat, sauteed onions, lettuce, tomatoes, pickled
gherkins, ketchup and mustard.

User: I don't like mustard

Chatbot(marked as "moderately appropriate"): Okay, this is
a serious topic, let me check. Ah, here it is: I contain
gluten and mustard.
```

Figure 5: Example for a chatbot response labeled as "moderately appropriate"

About half of the interactions (71) resulted in an appropriate and satisfactory response. Interestingly, triggering the fallback intent did not necessarily disrupted the conversation, as the fallback provided the user with three additional (random) options that suggested ideas for triggering intents to continue interacting with the chatbot.

Figure 6 shows the comparison between intended and matched intents for the most common user queries to our chatbots (excluding the fallback intent). While there were minimal issues with simple "welcome" intents, it is apparent that users often intended to express gratitude with a short "thank you" message, which falls under the category of "praise", which did not exist at the time of the study. This often resulted in highly inappropriate responses because we had not yet handled this intent.

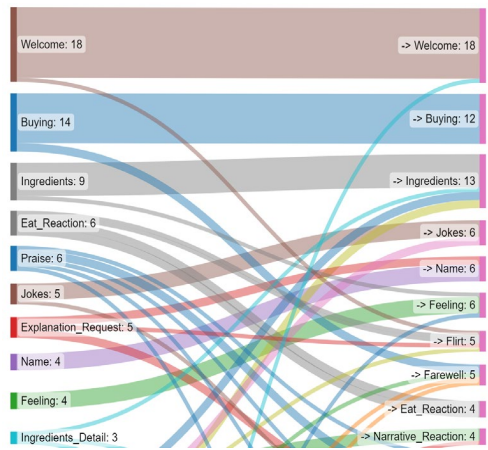

Figure 6: Intended (left side) and matched (right side) intents, of the 10 most frequently matched intents of our chatbots within the study

Next, we focused on the 145 user inquiries and mapped them to their targeted intents. In cases where an intent did not exist, we created a new intent. Each user inquiry was then assigned to one of three broader interaction categories based on the four categories of interaction with CAs defined by Weber and Ludwig (2020). The "entertainment" category accounted for the most interactions (87) and represented a tendency to anthropomorphize the chatbot (e.g., greeting the chatbot; expecting an appropriate response when telling to eat the food soon; praising it; asking it for jokes, names, or feelings). The second most common form of interaction (54 inquiries) was categorized as "information & advice", which combines the two categories "information requests" and "advice services" into one (e.g., asking about the ingredients of the meal; asking for more details about individual ingredients; asking about the preparation of the meal; or asking where to buy/order the meal). The last category, with only four queries, was the "control" category, which is derived from the "voice control" category (Weber and Ludwig, 2020), indicating that users attempted to trigger specific actions through the chatbot (e.g., using a "scan" prompt to try to activate the AR feature; asking for assistance in using the chatbot; or confirming a suggested intent option in a fallback response with "yes, please").

When examining interactions with the chatbot across the four phases (Figure 7), it is noteworthy that most interactions occurred during the *Prearrival* and *Postarrival & Preprocess* phases, with only one interaction *While Dining* and a slight increase in interactions *After Dining*.

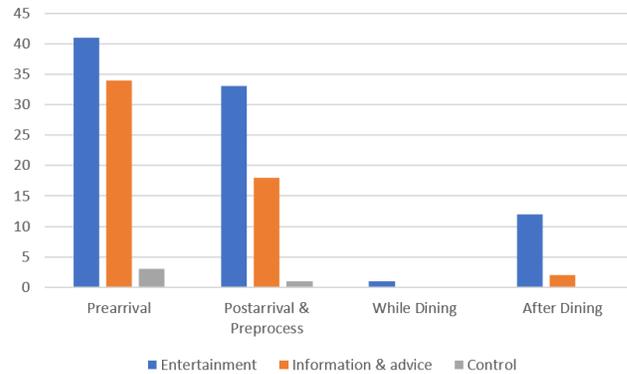

Figure 7: Aggregated user intentions with the chatbots for the four phases of eating out.

*6.2.6 Input and Output Modalities*

The participants in our study had different perceptions and preferences regarding the input and output modalities of the chatbots. While the speech-to-text functionality was recognized by G01, G03, and G05, only G01 used it. However, G01 expressed reluctance to use this feature in a restaurant, citing concerns about others' perceptions. The other participants had various reasons for not using speech-to-text, such as discomfort in the presence of others (G03, G06) or questioning its additional purpose (G07).

The speech output of the chatbot, which was only activated during AR interactions with the chatbot, was noticed by G01, G02, and G08. However, G03 (due to ambient noise) and G09 (due to a very quiet smartphone) did not perceive the speech output despite using the AR feature. Five participants (G01, G02, G03, G06, G09) expressed dislike of the speech output in restaurants or indicated that they would turn down the volume. Participants had different reasons for their preferences regarding speech output. Some expressed discomfort in the presence of others (G06), fear of negative perceptions (G01), or perceived it as an unnecessary distraction (G02). Avoiding attention was a concern for G03. On the other hand, G01 and G09 found the robotic pronunciation amusing, and G08 stated that he felt comfortable talking to the chatbot.

*6.2.7 Using the App in Different Scenarios*

When asked about their willingness to use the app in different scenarios, participants were generally willing to consider using the app in the future, either alone or with friends, and recognized its potential. However, they were less convinced about using it with strangers or people they barely know.

In scenarios where participants would eat out with others, four participants (G01, G03, G05, G07) considered the possibility that using the app could lead to different restaurant choices within the group, making the decision-making process more difficult than before, especially if everyone used the app individually on their own phones. Conversely, G08 and G09 had a different perspective and expressed their willingness to use the app on a single device. They believed that using the app together would increase the "fun factor" and lead to a more enjoyable experience. The reasons for not using the app with strangers were mainly related to the novelty of the app and the participants' unfamiliarity with it (G01, G04, G06, G09). According to G07, in scenarios with strangers, the focus should be on getting to know the other persons, and the presence of an app could hinder this process. However, G08 acknowledged the potential of the app as a conversation starter in such situations.

Regardless of group constellations, G02, G03, G07, and G08 expressed a greater tendency to use the app in unfamiliar cities than in familiar environments. This preference was due to their lack of knowledge about good restaurants in unfamiliar cities. For G03 and G05, the app also provided a useful orientation guide by displaying nearby meals and dining options with corresponding pictures and prices, eliminating the need for "cumbersome" searches on platforms such as Google and Google Maps (G05).

## 6.3 Evaluation II: Web Interface Setup and Applied Methods

The evaluation of the web interface for content creation involved four sessions with restaurant owners. The sessions took place in March 2022, during which we had the opportunity to work again with R1, R2, R5 (from our first user study; Table 2), and R8, the co-owner of R5. By involving many of the same participants, we wanted to assess whether the initial issues and challenges identified in our user study had been addressed in our implementation.

Sessions lasted between 37 and 51 minutes, with an average of 47 minutes. Two researchers were present during all sessions. After a brief explanation of the procedure and tasks, the researchers assumed a monitoring role and refrained from actively intervening in the participants' use of the web interface. They took unstructured observation notes to document the participants' interactions.

Participants were given the freedom to express their thoughts freely using a thinking aloud approach (Nielsen, 1994). They were given tasks such as adding and setting up their restaurant (e.g., opening hours), adding a meal to the platform, creating a virtual avatar for their meal, and exploring the dashboard for key performance indicators. At the end of the session, a brief semi-structured interview was conducted to inquire about specific moments, features, perceived strengths and weaknesses, and suggestions for improvement or additional features. The sessions were audio and screen recorded for further analysis.

Following the sessions, the two researchers discussed and compared their main observations and findings. The handwritten observation notes were digitized and the audio recordings were fully transcribed. This allowed for a careful examination of the material to ensure that no details were overlooked, and any unclear notes were clarified to ensure proper interpretation. The data analysis process was of deductive nature, using the interview questions and the implemented features of the web interface to structure the data.

## 6.4 Results of the Evaluation Study (Web Interface)

Overall, the web interface received positive feedback for its well-structured and intuitive design, making it easy for most participants to use and operate. Restaurant owners R5 and R8 particularly praised the ease of use, noting that they did not have to search long to find the features they wanted. However, R2 expressed that the interface felt overly complicated when it came to managing dishes.

For the chatbot approach, content management of the chatbots was enabled indirectly through our system, such as through the selection options when creating a dish (e.g., gender, ingredients). This eliminated the need for restaurateurs to write their own responses for the chatbots, which R1 and R2 appreciated. They mentioned that they probably would not have the time to create the responses themselves, or that they did not think they were creative enough for such a task. R1 also suggested that the style of the food avatars could be made more cartoonish. The location-based aspect of the chatbots received positive feedback, as it allowed restaurateurs to automatically send location-based notifications to app users highlighting their offerings. This feature was praised by R1 for its potential to attract customers. The dashboard and key performance indicators (KPIs) received mixed feedback. R5 found the graphics and language in the dashboard view complex, while R8 found the insights provided interesting.

Several potential improvements and additional features were identified during the evaluation study. One suggestion from R5 was the ability to place orders from within the application. Integration with existing point-of-sale systems was also recommended. R1 suggested that the food avatars could alert customers to discounted items that need to be sold before closing time to prevent spoilage. In addition, R1 suggested that the avatars could notify users in advance of parties or events taking place at his restaurant.

## 7 DISCUSSION

In our discussion, we want to shed light on the remaining design spaces, learning opportunities, and recommendations based on the insights we gained through the evaluations of the Rendezfood platform. Furthermore, our study hints at implications for "interactive points of interest" as a starting point for experiences in a real-world metaverse.

### 7.1 Open Design Spaces, Learning Opportunities, and Recommendations for the Rendezfood platform

This chapter explores specific enhancements and recommendations for the Rendezfood platform, addressing various aspects of UX, AR functionality, customization, and social interaction that could be incorporated into similar platforms for a more engaging and seamless dining experience. To improve the user experience, G03 and G05 suggested the inclusion of an overlay tutorial (Chou, 2015) that introduces users to key features of the application. The need for a tutorial or a more visible menu button is particularly felt when it comes to understanding the QR codes and AR functionalities.

In exploring alternative ideas for interactions with EAVAs (Weber et al., 2021), G09 suggested printing QR codes on burger buns instead of using QR code sticks to position the AR assets in the right place. Ideally, a system would not need position markers to correctly place the AR assets on the food. Additionally, users expressed a desire for responsive behavior from chatbots, such as reminding them when their food is getting cold (G01). Expanding on these ideas, EAVAs could dynamically adjust their behavior based on various contextual parameters, including weather, season, user sentiment, and the state of the food. With the current development of generative AI such as ChatGPT and the ability to easily create customized GPTs, this could be very interesting in the future.

Other key issues raised by users include fine-tuning their preferences and profiles for a better experience. G02 and G07 expressed the need for more specific selection options for cuisines. Users also wanted to choose between cold and hot meals, as well as breakfast, lunch, dinner, or dessert (G02, G04, G09). In addition, there were requests for filters based on nutritional values, including calorie information (G03, G04, G05, G07), which was originally in one of the UX stories but not implemented due to restaurateurs' concerns about the unavailability of this data. Some users also wanted to set a floor budget in addition to a maximum limit for their meal recommendation (G06, G07). The importance of menu cards and the desire to view restaurant menus within the app was emphasized by G02, G03, G06, G08, and G09, with G09 specifically highlighting the value of receiving special offers. To address these needs, the location-based approach of the *Rendezfood* app could be used to automatically display the full menu or a list of discounted specials to the user when they enter a restaurant.

Enhancing the social dimension of the dining experience is a key aspect that emerged from user feedback. To simplify the process of choosing a place to eat, users (G01, G02, G03, G07, G08) expressed interest in receiving recommendations from family, friends, and even strangers based on their previous dining experiences. The ability to combine multiple profiles within a group to find suitable dining options was suggested by G06. The inclusion of a chat feature was seen as beneficial for group decision making (G04). Furthermore, exploring group interactions with EAVAs beyond individual sessions could provide insights into various interpersonal aspects in a social context. In addition, the introduction of more playful elements and rewards was suggested by G06 and G07 as a way to add playfulness to the app. Including mini-games and offering discounts as a reward for frequent use were seen as desirable.

### 7.2   Design Implications for the Future of Eating Out

In Figure 8, we have outlined design dimensions/goals, and promising technologies as design implications for the future of eating out, based on the work presented in the related work chapter and the lessons learned from our studies. The figure also allows existing work to be placed and can be used to further develop new concepts and ideas.

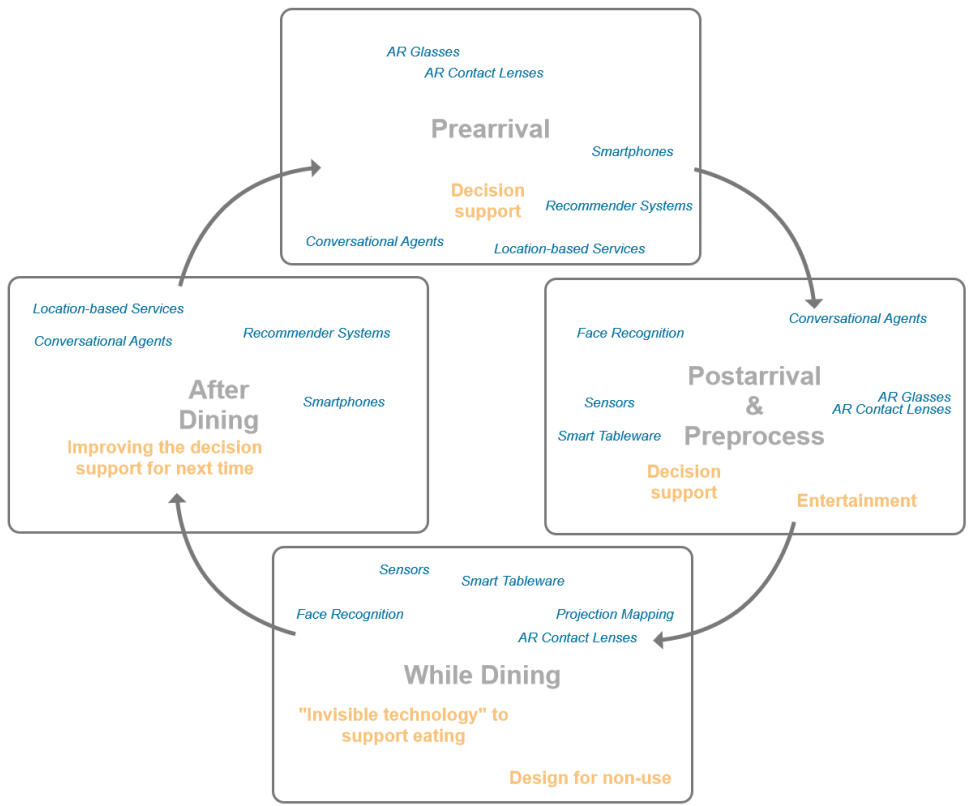

Figure 8: Design dimensions/goals (orange) and (potential) use of technology (blue) for the four phases of eating out

Examples of systems for the *Prearrival* phase include Lunchocracy (Terzimehić et al., 2018), which demonstrates chatbot-based decision making for group restaurant selection, or AllergyBot (Hsu et al., 2017), a chatbot-based system that provides restaurant recommendations based on individual allergies. An application such as *Rendezfood* is notable for its location-based and partially chatbot-based approach to recommending suitable meals before users decide on a restaurant. Its principle of "recommending meals, not restaurants" has been well received, and appreciated by users. It sets Rendezfood apart from traditional restaurant-focused platforms by placing a strong emphasis on the gastronomic experience. It is a commitment to prioritize the essence of dining itself, and it meets the aspirations of those seeking culinary adventure and new taste sensations, thus distinguishing Rendezfood from more traditional ad-focused platforms. Furthermore, *Rendezfood* not only facilitates *Prearrival*, but also incorporates entertainment elements to enhance the *Postarrival & Preprocessing* experience.

Another example is Le Petit Chef (Batat, 2021), which can be placed in the *Postarrival & Preprocess* phase, an AR application that uses projection mapping on the table to reduce waiting times through humorous animations. However, by incorporating CAs or other interactive elements, one could further increase the interactivity and thus the entertainment factor. The challenge then is to design the modalities in a way that does not disturb other visitors, for example, if the ICT artifact requires audio to be played. In addition, we see potential in AR menu cards (Arioputra and Lin, 2015; Lee et al., 2020; Obiorah et al., 2021), which could also be combined with CAs to support the decision-making process. Finally, the concept of smart coasters (Rocha et al., 2011), an example of invisible technology, has significant potential to subtly enhance the dining experience *While Dining* in several ways.

Despite these examples, we are still in search of an application that can accompany and support individuals in all four phases of their dining experience. While we indirectly tried to build such an application with Rendezfood, we discovered limitations during the evaluation. The virtual food avatars proved to be too intrusive during the dining phase, and not enough attention was paid to keeping users engaged with the system after the dining phase. We also found that the maintenance effort required for restaurant owners to enter and update their content was rather high, raising concerns about its feasibility. Novel approaches from the field of End-User Development (EUD) can remedy this, particularly with regard to Metaverse experiences (Krings et al., 2023). Recent advances in large language models (LLMs), conversational user interfaces (CUIs), and generative AI also offer new possibilities to challenge this issue. Generative AI and LLMs could greatly simplify content creation for restaurateurs, potentially enabling (semi-)automatic content generation. In addition, there is a growing interest among the general public to interact through CUIs, as evident by the high use of and interest in ChatGPT (Dwivedi et al., 2023).

### 7.3 Beyond Food: Interactive Points of Interest as a Central Starting Point for the Real-World Metaverse

By abstracting our concept from the food domain, the potential of interactive Points of Interest (iPOI) as a foundational element applicable to various contexts became apparent during our study. We define iPOI as interactive, location-based points embedded in the physical world that allow individuals or communities to easily create multi-modal content. These points can be interacted with by users of the platform when they are nearby.

The iPOI demonstrated in our paper includes the ability to easily position them on a GPS map, generate AR content (further detailed in Krings et al. (2022)), and customize the associated chatbots, even adjusting their audio output based on selected gender. While the creation of chatbot content is currently limited and specific to the food domain in our implementation, it is conceivable that a more generalized system could lead to engaging outcomes and opportunities in areas such as retail, tourism, career exploration, and emergency scenarios. This is especially compelling when all of these contextual domains are brought together in one unified platform that streamlines content creation for iPOI, including editing CA behaviors, adding audio files, and AR content (for virtual agents), which we understand to be a real-world metaverse, as recently brought up (Hanke, 2021; Xu, 2022). Therefore, we believe our work represents an important step forward in making the abstract concepts of the metaverse (Lee et al., 2021) and the real-world metaverse (Hanke, 2021; Xu, 2022) more tangible. By illustrating the potential of our approach in the food domain and extending it to iPOI, we contribute to the understanding and realization of these concepts.

### 8 LIMITATIONS AND FUTURE WORK

Throughout our research, we encountered certain limitations and potential for future work that need to be acknowledged for future studies and implementations. These aspects include primary difficulties with location-based notifications and considerations regarding the context of chatbot interactions. Testing the effectiveness of location-based notifications for chatbots proved challenging during our evaluation. Due to technical issues, notifications did not work properly for two participants, but the concept was still understood when explained in the interview.

Another consideration relates to the context of chatbot interactions. Our studies were conducted before the widespread use of advanced conversational AI models, such as ChatGPT. As a possible result, user queries were typically short, consisting of single-word requests such as "location", "directions", or "ingredients". It is worth considering how users may interact with chatbots and a system like *Rendezfood* differently today, potentially providing more detailed input and anthropomorphizing the chatbot to a greater extent (Wang et al., 2023) Additionally, users may expect the chatbot to retain information from previous messages and maintain the context of the conversation (Wang et al., 2023) which was only partially achieved in our implementation. These factors should be considered as they may affect the reproducibility of our findings in current or future studies. It should also be noted that all our evaluation studies were conducted under moderate weather conditions, and therefore the influence of extreme weather conditions (e.g., heat or heavy rain) on our intervention could not be observed, which is important to note as it can affect the practice of eating out (Wenzer, 2010).

With our overview of the use of technologies in the phases of going out to eat, we also initially want to create a structure for researchers. Our model does not yet incorporate, for instance, that design goals might be differentiated by the strong links between culture and eating out, or by the different locations and contexts of restaurants and venues (e.g., there are different design opportunities around an ice cream parlor compared to a fast-food restaurant or a steakhouse, or even food trucks (Adibah et al., 2018)). In future research, building on our results, these context-dependent differences could be explored.

## 9 CONCLUSION

The restaurant industry has faced many challenges in recent years, including the impact of COVID-19, the rise of delivery, inflation, and low profit margins. Despite these challenges, there is often untapped potential in leveraging technology for process optimization and customer engagement. In our design case study, we explored innovative approaches to improving the customer experience in restaurants and examined the technologies that are already being used for this purpose. Through our research, we identified technology needs of restaurateurs and reported on the needs of guests in the restaurant environment. We found that there is great potential for location-based systems and CAs to enhance the eating out experience. However, we also discovered the challenges associated with creating content for such platforms. These findings contribute to the field of HFI by providing insights into customer requirements, presenting an innovative approach to improving the user experience, and exploring promising future directions based on our design case study findings.

Our work revealed that existing interventions in the practice of eating out often focus on isolated phases, with limited attention paid to the *While Dining* and *After Dining* phases. We designed an ICT artifact, based on the concept of EAVAs (Weber et al., 2021), that was made available in restaurants in a location-based manner and customizable by restaurant owners. Evaluation of the prototype provided valuable insights into the design of socially acceptable technological interventions in this context and identified promising avenues for future research. Our research also revealed design implications for digital platforms aimed at enhancing the dining out experience. We found that interactive features such as personalized menu recommendations and menu item customization create a more engaging and personalized experience for customers. In addition, location-based marketing services can help address the challenge of effectively promoting restaurants to the local community.

Transparency and responsiveness to events or circumstances were highlighted as essential factors for restaurant owners in their marketing efforts.

Looking to the future, we envision the development of a holistic solution that supports individuals through all four phases of their dining experience and beyond. Although our own prototype, *Rendezfood*, had limitations during evaluation, recent advances in large-scale language models, conversational user interfaces, and generative AI offer new possibilities. These advances can simplify content creation for restaurant owners and drive greater user engagement. Our work lays the foundation for a real-world metaverse and may influence its development. We also recognized the potential of interactive points of interest (iPOI) as a foundational element of a real-world metaverse applicable to various contexts beyond food. iPOI are interactive, location-based points that allow users to create content through an end-user development interface. This concept holds promise in retail, tourism, emergency scenarios and several other contexts. By abstracting our concept from the food domain, we contribute to the understanding and realization of the (real-world) metaverse and demonstrate the potential of iPOI as a central starting point for its implementation.

In conclusion, our research provides valuable insights into improving the customer experience in restaurants through innovative technological interventions. Researchers and practitioners can build on our findings to shape their own visions of the future, creating interactive and immersive experiences that enhance various aspects of our daily lives.

**ACKNOWLEDGMENTS**

This work was carried out as part of the Rendezfood project (ERDF-0801425), funded by the European Regional Development Fund. We would like to thank Sabrina Brodesser, Nora Hille, Laura Grönewald, Fady Tawfik, Daniel Benfer, Jörg Muschiol and Kai Gutberlet for their contributions to this work. We would also like to thank all the participants who took part in our studies.

Pieska, S., Luimula, M., Jauhiainen, J., Spiz, V., 2013. Social Service Robots in Wellness and Restaurant Applications. Journal of Communication and Computer 10, 116–123.

Quesenbery, W., Brooks, K., 2010. Storytelling for User experience: Crafting Stories for Better Design. Rosenfeld Media, New York.

Rocha, L., Lohmann, A., Braz, A., Bitarello, B., Reiszel, F., 2011. Smart Beer Coaster, in: IADIS International Conference on Interfaces and Human Computer Interaction 2011. pp. 384–388.

Sardella, N., Biancalana, C., Micarelli, A., Sansonetti, G., 2019. An Approach to Conversational Recommendation of Restaurants, in: Stephanidis, C. (Ed.), HCI International 2019 - Posters. HCII 2019. Communications in Computer and Information Science, Vol 1034. Springer, Cham, pp. 123–130. https://doi.org/10.1007/978-3-030-23525-3_16

Schaarschmidt, M., Höber, B., 2017. Digital booking services: comparing online with phone reservation services. Journal of Services Marketing 31, 704–719. https://doi.org/10.1108/JSM-04-2016-0145

Seo, K.H., Lee, J.H., 2021. The Emergence of Service Robots at Restaurants: Integrating Trust, Perceived Risk, and Satisfaction. Sustainability 13, 4431. https://doi.org/10.3390/su13084431

Shengzhi, W., 2015. Foodies: An augmented reality food ordering system [WWW Document]. URL http://www.wushengzhi.xyz/project01 (accessed 2.9.24).

Spence, C., 2020. Multisensory Flavour Perception: Blending, Mixing, Fusion, and Pairing within and between the Senses. Foods 9, 407. https://doi.org/10.3390/foods9040407

Spence, C., Mancini, M., Huisman, G., 2019. Digital Commensality: Eating and Drinking in the Company of Technology. Front Psychol 10. https://doi.org/10.3389/fpsyg.2019.02252

Steiniger, S., Neun, M., Edwardes, A., 2006. Foundations of Location Based Services. Lecture Notes on LBS 1, 1–28.

Sterckx, F., Verbeeck, A., 2023. Le Petit Chef [WWW Document]. Skullmapping. URL https://skullmapping.com/project/le-petit-chef/ (accessed 2.9.24).

Susskind, A.M., Curry, B., 2019. The Influence of Tabletop Technology in Full-Service Restaurants, in: Susskind, A., Maynard, M. (Eds.), The Next Frontier of Restaurant Management: Harnessing Data to Improve Guest Service and Enhance the Employee Experience. Cornell University Press, Ithaca, NY, pp. 203–214. https://doi.org/10.7591/9781501736520-016/HTML

Tan, T.F., Netessine, S., 2020. At Your Service on the Table: Impact of Tabletop Technology on Restaurant Performance. Manage Sci 66, 4496–4515. https://doi.org/10.1287/mnsc.2019.3430

Terzimehić, N., Khamis, M., Bemmann, F., Hussmann, H., 2018. Lunchocracy: Improving Eating Dynamics in the Workplace Using a Bot-Based Anonymous Voting System, in: Extended Abstracts of the 2018 CHI Conference on Human Factors in Computing Systems. ACM, New York, NY, USA, pp. 1–6. https://doi.org/10.1145/3170427.3188626

Wang, T., Wang, D., Li, B., Ma, J., Pang, X.S., Wang, P., 2023. The Impact of Anthropomorphism on Chatgpt Actual Use: Roles of Interactivity, Perceived Enjoyment, and Extraversion. (Preprint). https://doi.org/http://dx.doi.org/10.2139/ssrn.4547430

Wang, Y., Li, Z., Jarvis, R., Khot, R.A., Mueller, F.F., 2019. iScream!: Towards the Design of Playful Gustosonic Experiences with Ice Cream Yan, in: Extended Abstracts of the 2019 CHI Conference on Human Factors in Computing Systems. ACM, New York, NY, USA, pp. 1–4. https://doi.org/10.1145/3290607.3313244

Warde, A., 2018. Changing Tastes? The Evolution of Dining Out in England. Gastronomica: The Journal of Critical Food Studies 18, 1–12. https://doi.org/10.1525/gfc.2018.18.4.1

Weber, P., Engelbutzeder, P., Ludwig, T., 2020. "Always on the Table": Revealing Smartphone Usages in everyday Eating Out Situations, in: Proceedings of the 11th Nordic Conference on Human-Computer Interaction: Shaping Experiences, Shaping Society. ACM, New York, NY, USA, pp. 1–13. https://doi.org/10.1145/3419249.3420150

Weber, P., Krings, K., Nießner, J., Brodesser, S., Ludwig, T., 2021. FoodChattAR: Exploring the Design Space of Edible Virtual Agents for Human-Food Interaction, in: Designing Interactive Systems Conference 2021. ACM, New York, NY, USA, pp. 638–650. https://doi.org/10.1145/3461778.3461998

Weber, P., Ludwig, T., 2020. (Non-)Interacting with Conversational Agents: Perceptions and Motivations of Using Chatbots and Voice Assistants, in: Proceedings of the Conference on Mensch Und Computer. ACM, New York, NY, USA, pp. 321–331. https://doi.org/10.1145/3404983.3405513

Weber, P., Ludwig, T., Michel, L.K., 2023. The role of technology use in food practices during the COVID-19 pandemic. Int J Gastron Food Sci 31. https://doi.org/10.1016/j.ijgfs.2023.100687

Wei, J., Wang, X., Peiris, R.L., Choi, Y., Martinez, X.R., Tache, R., Koh, J.T.K.V., Halupka, V., Cheok, A.D., 2011. CoDine: An Interactive Multi-sensory System for Remote Dining, in: Proceedings of the 13th International Conference on Ubiquitous Computing. ACM, New York, NY, USA, pp. 21–30. https://doi.org/10.1145/2030112.2030116

Wenzer, J., 2010. Eating Out Practices Among Swedish Youth: Gothenburg Area Foodscapes, rapport nr.: CFK-rapport 2010: 03. Centrum för konsumentvetenskap (CFK), Handelshögskolan vid Göteborgs universitet.

Wulf, V., Rohde, M., Pipek, V., Stevens, G., 2011. Engaging with Practices: Design Case Studies as a Research Framework in CSCW, in: Proceedings of the ACM 2011 Conference on Computer Supported Cooperative Work - CSCW '11. ACM Press, New York, New York, USA, pp. 505–512. https://doi.org/10.1145/1958824.1958902

Xu, J., 2022. From Augmented Reality Location-based Games To the Real-world Metaverse, in: Extended Abstracts of the Annual Symposium on Computer-Human Interaction in Play. ACM, New York, NY, USA, pp. 364–366. https://doi.org/10.1145/3505270.3558363